\documentclass[aps,pra,twocolumn,showpacs,superscriptaddress]{revtex4}

\usepackage{graphicx}
\usepackage{amsmath}
\usepackage{amssymb}

\input{epsf}
\begin{document}

\title{Small mass- and trap-imbalanced two-component Fermi systems}

\author{D. Blume}
\affiliation{Department of Physics and Astronomy,
Washington State University,
  Pullman, Washington 99164-2814}

\date{\today}

\begin{abstract}
Motivated by the
prospect of 
optical lattice experiments with two-component Fermi gases consisting
of different atomic species such as Li and K,  
we calculate the energies for $N$ fermions under harmonic
confinement as a function of the mass- and trap-imbalance,
i.e., as a function of the ratio between the masses and frequencies 
of species one and two, using microscopic approaches.
Our energies for $N=2$ through $6$
can be used to determine the energetically most favorable
configuration for a given number of atoms per species
of a deep lattice in which each lattice site is approximately
harmonic and in which tunneling between neighboring sites
can be neglected.
Furthermore, our energies determine one of the input parameters, namely the
onsite interaction strength, of the corresponding lattice Hamiltonian.
We also determine and interpret 
the excitation gap for unequal-mass systems with up to 
$N=13$ atoms
for equal oscillator lengths.
\end{abstract}


\maketitle

\section{Introduction}
\label{introduction}
Cold-atom experiments have reached an impressive level of sophistication
over the past decade. About ten different atomic species have been Bose condensed
and, although experimentally more challenging, an increasing number
of
fermionic species have been cooled to quantum degeneracy, including
$^3$He~\cite{mcna06}, $^6$Li~\cite{trus01}, 
$^{40}$K~\cite{dema99} and two Yb isotopes 
($^{171}$Yb and $^{173}$Yb)~\cite{fuku07,fuku07a}. 
To date,
experiments on fermionic atoms have
focused on studying Bose-Fermi mixtures~\cite{modu02,gold02a,modu03,ospe06},
one-component Fermi systems with $p$-wave interactions~\cite{gaeb07}
and equal-mass two-component Fermi systems with
interspecies $s$-wave interactions~\cite{grei03,zwie03,stre03}.

Presently, the simultaneous
trapping and cooling of two different fermionic species is
being actively pursued by a number of laboratories~\cite{tagl06,tagl08,grimm}, 
adding a new
degree of freedom, i.e., the mass ratio between the
two atomic species.
Unequal-mass two-component Fermi systems 
are expected to behave quite differently than the equal-mass 
counterpart~\cite{efim71,efim73,petr05,kart07,wu06,iski06,lin06,orso07,pari07,iski08a}.
From the few-body perspective, the existence of 
weakly-bound trimers for sufficiently
large mass ratios 
consisting of two heavy fermions
and one light fermion 
is intriguing~\cite{efim71,efim73,petr05,kart07}.
Whether the existence of these bound trimer 
states allows, e.g., for the formation of a gas consisting of trimers
with sufficiently long lifetime has been
discussed~\cite{nish08}.
On the other hand,
adopting a many-body perspective~\cite{wu06,iski06,lin06,iski08a},
the ground state phase diagram of mass-
and population-imbalanced two-component Fermi
systems has been predicted
to show
quantum and topological phase transitions
which are not present in 
the phase diagram of population-balanced equal-mass
two-component Fermi systems.

The 
increasing interest of not only the atomic physics community
but also the nuclear physics, molecular physics,
condensed matter physics and quantum information 
science communities in cold atom systems
can be attributed to two major achievements.
First,
the atom-atom scattering length can be adjusted experimentally
to essentially any value, including vanishingly small 
and infinitely large positive or negative values,
by applying an external field in the vicinity of a so-called Fano-Feshbach 
resonance~\cite{stwa76a,ties93,inou98,corn00}.
Second, 
cold atomic gases can be loaded into an optical
lattice~\cite{grei02,stof04,koeh05,rom06}, 
allowing, e.g., for the study of the Mott-insulator 
transition~\cite{grei02},
a topic historically primarily considered by condensed matter 
physicists.
Furthermore, cold atom systems loaded into optical lattices may
ultimately be used as a 
quantum simulator~\cite{feyn82,lloy96,jane03}.

To date, most microscopic studies of equal-mass systems have assumed equal
trapping potentials of the two 
species~\cite{wern06b,jaur07,kest07,stec07c,chan07,blum07,stec07b,stec08}. 
However,
the lattice potential
felt by the two different hyperfine states may be different
even 
for equal-mass systems, 
leading to trap-imbalanced systems~\cite{iski08}. 
For unequal-mass systems such  as a 
$^6$Li-$^{40}$K
mixture (for which the mass ratio $\kappa$ is approximately $6.7$),
the trapping potentials felt by the two species are,
in general, different, owing to the mass difference and the 
species-dependent properties of the hyperfine
states. 
The trapping potentials felt by the two species may be tuned 
to some degree experimentally~\cite{iski08,orso07}.
Motivated by these considerations,
the present paper explores the rich behavior of trap- and
mass-imbalanced systems. These systems
share some
similarities with 
population-imbalanced systems~\cite{zwie06,part06}, 
which have received considerable
attention recently.

Assuming a deep lattice with
neglegibly small tunneling between neighboring lattice sites,
this paper determines the ground state properties of small
$s$-wave interacting 
two-component Fermi
systems trapped by spherically symmetric harmonic species-specific
potentials with trapping frequencies $\omega_1$ and $\omega_2$,
respectively.
Throughout, we adopt a microscopic many-body framework.
Our main results are:
(i) For small but
negative $s$-wave scattering lengths, 
we determine a compact expression for 
the ground state energy of small systems
with unequal masses and trapping frequencies 
perturbatively.
(ii) 
In the strongly-interacting unitary regime,
our numerical energies 
determine the phase diagram of optical lattice systems
in the no-tunneling regime for a large range
of mass ratios and trapping frequencies and
the on-site interaction strengths that parametrize the corresponding
lattice Hamiltonian; furthermore, they provide
insights into the behavior of the excitation gap.
(iii) We show explicitly that the behavior of trap-imbalanced
systems with small and
positive $s$-wave scattering lengths is---just as that of trap-balanced
systems---to
leading order governed
by the dimer-dimer scattering length.

Section~\ref{sec_system}
introduces the Hamiltonian and the numerical
techniques employed to solve the corresponding time-independent
Schr\"odinger equation.
Section~\ref{sec_results} contains our results for
small negative, infinitely large and small positive $s$-wave
scattering lengths.
Finally, Sec.~\ref{sec_conclusion} concludes.

\section{Theoretical background}
\label{sec_system}

\subsection{Hamiltonian}
\label{sec_ham}
The adopted model Hamiltonian $H$ for a
two-component Fermi system with $N_1$ mass $m_1$ and $N_2$ mass $m_2$
atoms
under spherically
harmonic confinement 
reads
\begin{eqnarray}
\label{eq_ham}
H = 
\sum_{i=1}^{N_1} \left(
\frac{-\hbar^2}{2m_1} \nabla^2_i + \frac{1}{2} m_1 \omega_{1}^2
\vec{r}_i^2 \right) + \nonumber \\
\sum_{i'=1}^{N_2} \left(
\frac{-\hbar^2}{2m_2}  \nabla^2_{i'} + 
\frac{1}{2} m_2 \omega_{2}^2
\vec{r}_{i'}^2 \right) + 
\sum_{i=1}^{N_1} \sum_{i'=1}^{N_2} V(\vec{r}_{ii'}),
\end{eqnarray}
where $\vec{r}_i$ and $\vec{r}_{i'}$ denote the position
vectors of the $i$th atom of species 1 and the $i'$th
atom of species 2, respectively, and $\omega_1$ and $\omega_2$ the
angular trapping frequencies felt by the atoms of species 1 and 2, 
respectively.
The interaction potential $V$ depends on the interparticle
distance vector $\vec{r}_{ii'}$, $\vec{r}_{ii'} =  \vec{r}_i - \vec{r}_{i'}$,
and is characterized by 
the $s$-wave scattering length $a_s$.
Throughout, like atoms are 
assumed to be
non-interacting, which is well justified for most experimentally
relevant systems.

Our perturbative, small $|a_s|$ analysis (see Sec.~\ref{sec_negative})
considers a zero-range $\delta$-function potential 
$V_{\delta}(\vec{r})$~\cite{ferm34},
\begin{eqnarray}
\label{eq_pp}
V_{\delta}(\vec{r}) = \frac{2 \pi \hbar^2 a_s}{\mu}  \delta(\vec{r}),
\end{eqnarray}
where $\mu$ denotes the reduced mass, 
$\mu = m_1 m_2 / (m_1+m_2)$.
In our numerical
calculations (see Secs.~\ref{sec_large} and \ref{sec_positive})
we employ,
as in our previous calculations~\cite{stec07b,blum07,stec08},
a shape-dependent spherically-symmetric
square well potential $V_{sw}(r)$
with range $R_0$ and depth $V_0$ ($V_0>0$),
\begin{eqnarray}
\label{eq_sw}
V_{sw}(r) = \left\{ \begin{array}{cl}
-V_0 & \mbox{for } r < R_0 \\
0 & \mbox{for } r > R_0
\end{array} \right. ,
\end{eqnarray}
where $r=|\vec{r}|$.
For a fixed $R_0$, $V_0$ is adjusted so that the interspecies 
$s$-wave scattering
length $a_s$ takes on the desired value.
Section~\ref{sec_large} considers the
so-called unitary regime, where
$V_0$ is adjusted so that the
two-body potential supports a zero-energy $s$-wave bound state,
implying a diverging $s$-wave scattering length $a_s$, i.e., $1/a_s=0$,
but no deeply-lying bound states.
Section~\ref{sec_positive}, in contrast, considers 
the regime where
$V_0$ is adjusted so
that the free-space dimer
supports one deep-lying $s$-wave bound state
(whose binding energy
depends on
the details of the two-body potential), implying a
small positive
$s$-wave scattering length.
The range $R_0$ of $V_{sw}$ is taken 
to be small compared to the
oscillator lengths $a_{ho,i}$,
\begin{eqnarray}
\label{eq_ho12}
a_{ho,i} = \sqrt{\hbar/(m_i\omega_i)},
\end{eqnarray}
where $i=1$ or 2.
Most calculations reported below use
$R_0=0.01 a_{ho,1}$.
To estimate how well the resulting properties agree
with those for zero-range interactions,
we analyze 
the dependence of the observables on the range $R_0$ in detail
for a few selected cases.

Section~\ref{sec_results} 
presents our results for three different scattering length
regimes, i.e., for weakly-attractive Fermi gases
($|a_s|$ small and $a_s<0$), for strongly-interacting Fermi gases
($1/|a_s| = 0$) and for
weakly-repulsive Fermi gases ($a_s$ small and $a_s>0$).
In all three regimes, we determine the energies of small trapped systems
with either $N_1=N_2$ or $|N_1-N_2|=1$. In addition to
changing the number of particles
and the scattering length $a_s$, we vary the mass ratio $\kappa$,
\begin{eqnarray}
\kappa = m_2/m_1,
\end{eqnarray}
and the ratio $\omega_2/\omega_1$ between the two trapping frequencies.
Mass ratios ranging from $\kappa=1$ to 8 are considered
(for unequal-mass systems species 2 has the heavier mass).
For $\kappa \gtrsim 8.6$,
three-body bound states have been
predicted to exist for systems
that consist
of two heavy fermions
and one light fermion and
that interact through zero-range potentials~\cite{petr05,kart07}. 
While studying the implications of these three-body states
for many-body systems
is interesting (see, e.g., Ref.~\cite{nish08}), this
topic is beyond the scope of the present paper.

\subsection{Numerical techniques}
\label{sec_num}
To solve the time-independent Schr\"odinger equation for the
Hamiltonian given in Eq.~(\ref{eq_ham}), we employ
two different numerical techniques. For $N_1=N_2=1$, we 
first build and then diagonalize
the Hamiltonian matrix while we resort to the 
fixed-node diffusion quantum Monte Carlo
(FN-DMC)
technique~\cite{hamm94,reyn82} for larger systems.

We first discuss the diagonalization approach 
employed to solve the Schr\"odinger equation for the
Hamiltonian given in Eq.~(\ref{eq_ham}) with $N_1=N_2=1$ and
$V=V_{sw}$; it follows Ref.~\cite{deur07},
with the main difference that we use a finite-range square-well
potential
while Ref.~\cite{deur07} uses the 
Fermi-Huang pseudo-potential~\cite{huan57}.
We rewrite our two-body Hamiltonian $H_{tb}$ in terms 
of a center-of-mass Hamiltonian $H_{cm}$, a 
relative Hamiltonian $H_{rel}$ and a coupling term 
$V_{coup}(\vec{R},\vec{r})$~\cite{deur07},
\begin{eqnarray}
\label{eq_hamtwobody}
H_{tb}  = H_{cm} + H_{rel} + V_{coup}(\vec{R},\vec{r}),
\end{eqnarray}
where
\begin{eqnarray}
\label{eq_hamcm}
H_{cm} = 
\frac{-\hbar^2}{2M} \nabla^2_{\vec{R}} + 
\frac{1}{2}M \omega_{cm}^2 R^2,
\end{eqnarray}
\begin{eqnarray}
\label{eq_hamrel}
H_{rel}=
\frac{-\hbar^2}{2 \mu} \nabla^2_{\vec{r}} +
\frac{1}{2} \mu \omega_{rel}^2 r^2,
\end{eqnarray}
and
\begin{eqnarray}
\label{eq_hamcoup}
V_{coup}(\vec{R},\vec{r}) = \mu \omega_{coup}^2  \vec{R} \cdot \vec{r}.
\end{eqnarray}
Here, $\vec{R}$ and $\vec{r}$ ($\vec{r} = \vec{r}_1-\vec{r}_2$)
denote the center-of-mass and relative vectors,
respectively, and $M$ denotes the total mass of the two-body system,
$M=m_1+m_2$.
The 
frequencies $\omega_{cm}$, $\omega_{rel}$ and $\omega_{coup}$
are defined as
\begin{eqnarray}
\omega_{cm} = \sqrt{(m_1 \omega_1^2 + m_2 \omega_2^2)/M},
\end{eqnarray}
\begin{eqnarray}
\omega_{rel} = \sqrt{(m_2 \omega_1^2 + m_1 \omega_2^2)/M}
\end{eqnarray}
and
\begin{eqnarray}
\omega_{coup} = \sqrt{|\omega_1^2 - \omega_2^2|}.
\end{eqnarray}

For equal trapping frequencies,
$V_{coup}$ vanishes and both
$\omega_{cm}$ and $\omega_{rel}$ reduce to $\omega_1$
(which equals $\omega_2$). In this case,
the center-of-mass and relative motions decouple, and the total
wave function $\Psi(\vec{R},\vec{r})$ 
can be written as a product of a $\vec{R}$-dependent
function $\Phi_{NLM_L}$
and a $\vec{r}$-dependent function $\phi_{nlm_l}$.
The $\Phi_{NLM_L}$ and $\phi_{nlm_l}$ 
are solutions to the Schr\"odinger equations for $H_{cm}$ and
$H_{rel}$, respectively,
and the subscripts
$NLM_L$ and $nlm_l$ denote the principal, angular momentum
and projection quantum numbers
of the center-of-mass and relative systems, respectively.
The $\Phi_{NLM_L}$ are the harmonic oscillator 
wave functions of a mass $M$ particle
with eigenenergies $E_{NL}$,
\begin{eqnarray}
E_{NL}=\left(2N + L + \frac{3}{2} \right) \hbar \omega_{cm},
\end{eqnarray} 
where $N=0,1,\cdots$, $L=0,1,2,\cdots$ and $M_L=-L,-L+1,\cdots,L$.
For the spherically-symmetric square well potential $V_{sw}$, 
the angular part of the relative wave function $\phi_{nlm_l}$
is given by the spherical harmonic $Y_{lm_l}$
while the radial part $R_{nl}$
can be written in terms of the confluent
hypergeometric function $M$ for $r<R_0$
and the Kummer function $U$ for $r>R_0$ (see, e.g.,
Ref.~\cite{kanj06}).
Equating the log-derivative of the inner and outer radial
wave functions
at $r=R_0$
results in a compact expression for the  
eigenequation,
from which we obtain the
eigenenergies $E_{nl}$ of $H_{rel}$
using standard root-finding techniques.
The radial wave functions $R_{nl}$
are then readily obtained by
enforcing continuity 
at $r=R_0$. We normalize the $R_{nl}(r)$ numerically.

To determine the eigenenergies of the two-particle Hamiltonian
$H_{tb}$ with 
non-zero $V_{coup}(\vec{R},\vec{r})$,
we expand the full wave function $\Psi(\vec{R},\vec{r})$ 
in terms of the complete set
$\{ \Phi_{NLM_L}(\vec{R}) \phi_{nlm_l}(\vec{r})\}$.
Recognizing that $H_{tb}$ commutes with the $z$-component of the
total angular momentum operator (i.e., that $M_L+m_l$ is 
conserved), we restrict the allowed
$M_L$ and $m_l$ combinations to $M_L+m_l=0$.
Since the $\Phi_{NLM_L}$ and $\phi_{nlm_l}$ are 
solutions of $H_{cm}$ 
and $H_{rel}$, respectively,
$H_{cm}$ and $H_{rel}$ are diagonal in this representation.
To evaluate the matrix elements involving $V_{coup}$,
we
rewrite the dot product $\vec{R} \cdot \vec{r}$ in terms of 
$R$, $r$, and the spherical harmonics $Y_{L=1,M_L}$ and $Y_{l=1,m_l}$
associated with 
the 
center-of-mass and relative degrees of freedom, respectively. 
The angular integrals then readily
reduce to Clebsch Gordon coefficients (multiplied by trivial 
constants), and the 
radial integrals are performed numerically.
The number of basis functions needed to converge the ground state energy
to a given relative accuracy strongly depends on the interaction
strength considered. At unitarity, e.g., we need a larger 
basis set than in the regime where $a_s$ is small and positive
(see Secs.~\ref{sec_large}
and \ref{sec_positive}).

The computational effort of diagonalization 
schemes such as that outlined above increases dramatically with increasing
number of particles, and eventually becomes
computationally unfeasible.
For larger number of particles, we thus
resort to an alternative numerical approach,
the FN-DMC method~\cite{hamm94,reyn82}, 
which exhibits a more favorable scaling with 
increasing number
of particles.
Our implementation of the 
FN-DMC method has been discussed in detail
in two recent papers~\cite{stec07b,stec08}; 
here, we only review the key points.

The FN-DMC technique, as used
throughout this paper, determines 
an approximate 
energy of the many-body system whose corresponding eigenfunction
has the same symmetry as a so-called guiding function $\psi_T$, i.e.,
the FN-DMC technique determines the energy of a
state that has the same nodal surface as $\psi_T$ but that may
differ from $\psi_T$ in other regions of the configuration space. 
If the nodal surface of $\psi_T$ coincides with that of the true 
eigenfunction, then the FN-DMC method results---within 
the statistical uncertainty that stems from the stochastic
nature of the approach---in the exact eigenenergy.
If the nodal surface of $\psi_T$ 
differs from that of the true eigenfunction, then
the FN-DMC method results in an upper bound to the true eigenenergy
whose eigenstate has the same symmetry as $\psi_T$. 
For example, $\psi_T$ can be constructed so as to obtain an upper
bound for the lowest eigen energy with total angular momentum
$L_{tot}=0$ or 1~\cite{stec08}.
In this paper, we restrict our FN-DMC calculations
to the energetically lowest-lying gas-like state of the system.

We consider three different 
parametrizations of the guiding function $\psi_T$:
(i) A guiding function $\psi_{T1}$ whose nodal surface is
constructed by anti-symmetrizing a pair function.
If $N$ is odd, a single-particle orbital is added (with 
the proper anti-symmetrization).
The detailed functional form of $\psi_{T1}$ 
is given by Eqs.~(35)-(39) of Ref.~\cite{stec08}.
(ii) A guiding function $\psi_{T2}$ whose nodal surface coincides with that
of the non-interacting ideal-gas nodal surface for the same number of
fermions of species 1 and species 2. The parametrization
follows that given by Eq.~(40) of Ref.~\cite{stec08}.
(iii) A guiding function $\psi_{T3}$ whose functional form allows,
at least in principle, to interpolate between the nodal 
surfaces of $\psi_{T1}$ and $\psi_{T2}$.
The functional form is given by Eqs.~(3)-(4) of Ref.~\cite{chan07}.

\section{Results}
\label{sec_results}

\subsection{Small negative $s$-wave scattering length}
\label{sec_negative}

This section considers the properties of small two-component 
Fermi systems with unequal masses and unequal trapping frequencies
in the
weakly-attractive regime, where $|a_s|$ is small ($a_s<0$).
In this regime,
a compact expression for the ground state
energy of
the Hamiltonian given in Eq.~(\ref{eq_ham})  
can be determined within first order degenerate perturbation theory
for the Fermi pseudo-potential $V_{\delta}$, Eq.~(\ref{eq_pp}).
Denoting the energy of the non-interacting system 
with $N_1$ atoms of mass $m_1$ and $N_2$ atoms of mass
$m_2$ by $E^{NI}_{N_1,N_2}$
(see Table~\ref{tab_pert} for selected values),
the perturbative expression for the energy $E_{N_1,N_2}$ 
reads
\begin{eqnarray}
\label{eq_pert}
E_{N_1,N_2} \approx 
E_{N_1,N_2}^{NI} + \hbar \bar{\omega} \frac{a_s}{\bar{a}_{ho}}
C_{N_1,N_2},
\end{eqnarray}
where
\begin{eqnarray}
\label{eq_wnatural}
\bar{\omega} = \frac{M\omega_1 \omega_2}{m_1 \omega_1 + m_2 \omega_2}
\end{eqnarray}
and 
\begin{eqnarray}
\label{eq_ahonatural}
\bar{a}_{ho} = \sqrt{\frac{\hbar}{2 \mu \bar{\omega}}}=
\sqrt{\frac{a_{ho,1}^2 + a_{ho,2}^2}{2}}.
\end{eqnarray}
The quantities $\bar{\omega}$ and $\bar{a}_{ho}$ have been defined
so that the coefficient $C_{1,1}$
is constant (i.e.,
independent of 
$\eta$, see below).
We refer to
$\bar{\omega}$ 
and
$\bar{a}_{ho}$ as the ``natural angular trapping frequency''
and the ``natural oscillator length''
of the two-body system in the BCS regime.

The coefficients $C_{N_1,N_2}$
are listed in Table~\ref{tab_pert} for selected
\begin{table}
\caption{
Energies $E^{NI}_{N_1,N_2}$
of the non-interacting 
system
and
dimensionless coefficients
$C_{N_1,N_2}$  
that determine the ground state
energy of
weakly-attractive trap- and mass-imbalanced
two-component
Fermi gases for selected $N_1$ and $N_2$ values.
The subscript pair $(j,k)$ can take the values $(1,2)$ or $(2,1)$.
}
\begin{ruledtabular}
\begin{tabular}{ll|c c}
  $N_j$ & $N_k$ & $E^{NI}_{N_1,N_2}/\hbar$ & $C_{N_j,N_k} \times (2 \sqrt{2 \pi}) $ \\
\hline
1 & 1 & $\frac{3}{2} (\omega_j+ \omega_k)$ & 4\\
2 & 1 &$4 \omega_j+\frac{3}{2} \omega_k$&  $(2 a_{ho,j}^2 + 4 a_{ho,k}^2)/\bar{a}_{ho}^2$ \\
2 & 2 &$4(\omega_j+ \omega_k)$& $(2 a_{ho,j}^4 + 9 a_{ho,j}^2 a_{ho,k}^2 + 2 a_{ho,k}^4)/\bar{a}_{ho}^4$ \\
3 & 2 &$\frac{13}{2} \omega_j+4 \omega_k$& $(2 a_{ho,j}^4 + 10 a_{ho,j}^2 a_{ho,k}^2 + 3 a_{ho,k}^4)/\bar{a}_{ho}^4$ \\
3 & 3 &$\frac{13}{2} (\omega_j+ \omega_k)$& $(3 a_{ho,j}^4 + 16 a_{ho,j}^2 a_{ho,k}^2 + 3 a_{ho,k}^4)/\bar{a}_{ho}^4$\\
4 & 3 &$9 \omega_j+\frac{13}{2} \omega_k$& $(3 a_{ho,j}^4 + 17 a_{ho,j}^2 a_{ho,k}^2 + 4 a_{ho,k}^4)/\bar{a}_{ho}^4$\\
4 & 4 &$9( \omega_j+ \omega_k)$& $(4 a_{ho,j}^4 + 23 a_{ho,j}^2 a_{ho,k}^2 + 4 a_{ho,k}^4)/\bar{a}_{ho}^4$\\
\end{tabular}
\end{ruledtabular}
\label{tab_pert}
\end{table}
$N_1$ and $N_2$ combinations with $|N_1-N_2|=0$ or 1 ($N \le 8$).
They reduce to those
reported in Ref.~\cite{stec08} for equal masses and equal frequencies.
The coefficients
$C_{N_1,N_2}$ in Table~\ref{tab_pert} are written
in terms of $a_{ho,1}$ and $a_{ho,2}$;
alternatively, they can 
be written in terms  of 
$\eta$, 
\begin{eqnarray}
\label{eq_eta}
\eta = 1- \left(\frac{a_{ho,2}}{a_{ho,1}} \right)^2.
\end{eqnarray}
The quantity $\eta$  
measures the density imbalance of the non-interacting two-component
Fermi gas.
For $\eta=0$, the oscillator lengths $a_{ho,i}$ ($i=1$ and 2) coincide;
for closed shell systems with $N_1=N_2$, this implies fully overlapping 
densities of the two non-interacting components.
For $\eta <0$, we have
$a_{ho,2}>a_{ho,1}$, while for $\eta>0$, we have 
$a_{ho,2}<a_{ho,1}$. 
For $\kappa=4$, e.g., $\eta < 0$ corresponds to
$\omega_2 < \omega_1/4$ and $\eta>0$ corresponds to $\omega_2 > \omega_1/4$.

\begin{figure}
\includegraphics[angle=270,width=90mm]{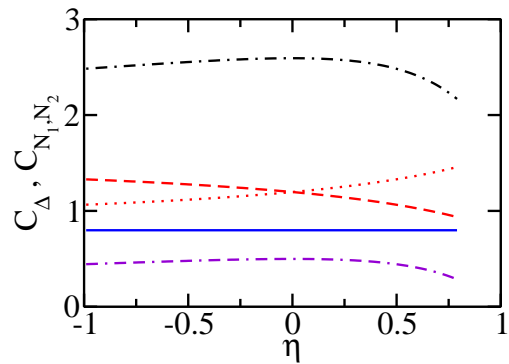}
\caption{
(Color online)
Dimensionless coefficients $C_{1,1}$
(solid line), $C_{2,1}$ (dashed line), $C_{1,2}$ (dotted line)
and $C_{2,2}$ (dash-dotted line) as a function of $\eta$
for weakly-attractive two-component Fermi gases.
The dash-dash-dotted line shows the dimensionless quantity 
$C_{\Delta}$, Eq.~(\ref{eq_cdelta}),
which determines
the excitation gap $\Delta(N)$ for $N=3$, $5$ and 7.
}
\label{fig_pert}
\end{figure}

Figure~\ref{fig_pert} shows the dimensionless
coefficients $C_{N_1,N_2}$ for $N \le 8$
as a function of $\eta$. Plotted this way, the coefficients
$C_{N_1,N_2}$ for fixed $N_1$, $N_2$ and $\eta$
but different $\kappa$ collapse to a single curve.
For $N=2$, $C_{N_1,N_2}$
is constant (see above). 
For $N=4$, $C_{N_1,N_2}$ is maximal for $\eta=0$ and decreases
as $|\eta|$ increases. This 
implies that the attractive interspecies scattering length
$a_s$ can most effectively
introduce correlations that lead to a lowering of the energy,
compared to $E_{N_1,N_2}^{NI}$,
when the densities of the two components
overlap fully.
For $N=3$, the coefficient $C_{2,1}$ decreases with increasing
$\eta$ while the coefficient $C_{1,2}$ increases with increasing $\eta$.
This ``asymmetry'' can be understood by realizing that 
the maximal density overlap of the two components for odd-$N$ systems
occurs for finite $\eta$
and not for $\eta=0$.
For $N_1=2$ and $N_2=1$, e.g., the maximal density 
overlap of the non-interacting system occurs for $\eta<0$;
consequently, the $C_{2,1}$ coefficient decreases with increasing $\eta$.
For $N_1=1$ and $N_2=2$, in contrast, the maximal density overlap
of the non-interacting system occurs for $\eta>0$.
This explains the reversed behavior of $C_{2,1}$ and $C_{1,2}$ as
a function of $\eta$.

The energies for systems with even and odd total number of atoms
determine the
excitation gap $\Delta(N)$~(see, e.g., Ref.~\cite{gior07}),
\begin{eqnarray}
\label{eq_gapgeneral}
\Delta(N) = \frac{E_{(N-1)/2,(N+1)/2}+E_{(N+1)/2,(N-1)/2}}{2}
- \nonumber \\ 
\frac{E_{(N-1)/2,(N-1)/2}+E_{(N+1)/2,(N+1)/2}}{2},
\end{eqnarray}
where we have taken $N$ to be odd.
Using the perturbative energy expression, Eq.~(\ref{eq_pert}), 
we find
\begin{eqnarray}
\label{eq_gap}
\Delta(N) \approx
 -\hbar \bar{\omega} \frac{a_s}{\bar{a}_{ho}} C_{\Delta},
\end{eqnarray}
where 
\begin{eqnarray}
\label{eq_cdelta}
C_{\Delta} = 
\frac{1}{4 \sqrt{2 \pi}} \;
\frac{5 a_{ho,1}^2 a_{ho,2}^2}{\bar{a}_{ho}^4}
\end{eqnarray}
for $N=3$, 5 and 7.
The excitation gap 
determined perturbatively 
is
independent of $N$
for $N \le 7$ for all $m_i$ and $\omega_i$ combinations.
A dash-dash-dotted line in Fig.~\ref{fig_pert} shows $C_{\Delta}$
as a function of
$\eta$.
The coefficient $C_{\Delta}$, and consequently also
$\Delta(N)$, is largest for $\eta=0$ and decreases with increasing
$|\eta|$.
This can be readily understood by realizing that the energies for
odd-$N$ systems
[first term on the right hand side of Eq.~(\ref{eq_gapgeneral})]
average to a constant, and that the average of the 
energies for even-$N$ system
[second term on the right hand side of Eq.~(\ref{eq_gapgeneral})]
is minimal for $\eta=0$.

The equal-frequency systems with mass ratio $\kappa$
correspond to 
$\eta = (\kappa-1)/\kappa$. 
Figure~\ref{fig_pert}
shows that the 
coefficient $C_{\Delta}$
decreases with increasing mass ratio
$\kappa$ for systems with $\omega_1=\omega_2$, in agreement with 
the findings of Ref.~\cite{stec08}.

\subsection{Infinitely large $s$-wave scattering length}
\label{sec_large}
This section considers 
infinitely strongly interacting two-component
Fermi systems with diverging $s$-wave scattering length
$a_s$ and varying mass and frequency ratios.
Throughout this section, we express energies in units
of the average oscillator energy $\hbar \omega$,
\begin{eqnarray}
\hbar \omega = \frac{\hbar \omega_1 + \hbar \omega_2}{2};
\end{eqnarray} 
this unit is convenient
since the energies of the non-interacting
systems with $N_1=N_2$ are directly proportional to
$\hbar \omega$
(see, e.g., Table~\ref{tab_pert} for small $N$).
For $N=3-14$ atoms, we determine the eigenenergies of the stationary
Schr\"odinger equation by the FN-DMC method. For $N=2$, we compare the
diffusion Monte Carlo (DMC) 
energies (in this case, the ground state of the system is nodeless
and no nodal approximation needs to be made) with the energies
obtained from the diagonalization scheme.

Table~\ref{tab_n2} reports selected two-body energies ($N_1=N_2=1$) for 
$\kappa=1$, 4 and 8 at unitarity.
The energies in the third and fourth column are calculated 
for the
square well potential with $R_0=0.01 a_{ho,1}$ 
using the DMC and the
diagonalization approaches, respectively.
We analyzed the convergence of the energies obtained by the diagonalization
approach by considering basis sets with up to about 1000
basis functions.
Within the
statistical uncertainties of the DMC energies,
the values reported in
columns three and four agree.
To estimate the energy's dependence 
on the range $R_0$ of the square well potential, 
we diagonalize the Hamiltonian
matrix for different
$R_0$.
We find that the energies for fixed 
ratio and frequency ratios vary linearly with $R_0$,
allowing for a simple linear extrapolation
to the  $R_0 \rightarrow 0$ limit
(see Ref.~\cite{stec08} for a similar analysis of equal-frequency systems).
The energies for $R_0=0.01 a_{ho,1}$
are slightly larger than the extrapolated zero-range energies
(seventh column of Table~\ref{tab_n2})
for all mass and frequency ratios
considered, and 
deviate by less
than 0.5\% from the
extrapolated zero-range energies.
Our extrapolated 
two-body energies at unitarity for equal frequencies 
equal 
$2 \hbar \omega$ for all $\kappa$, in agreement with analytical
results for the zero-range
potential~\cite{busc98}. 
For larger systems (see below)
we do not
explicitly extrapolate to the zero-range limit. Based on our two-body
results, we estimate
that the finite range effects of the FN-DMC energies for the larger systems
at unitarity are at most about a few times larger
than the statistical uncertainties.

\begin{figure}
\includegraphics[angle=270,width=90mm]{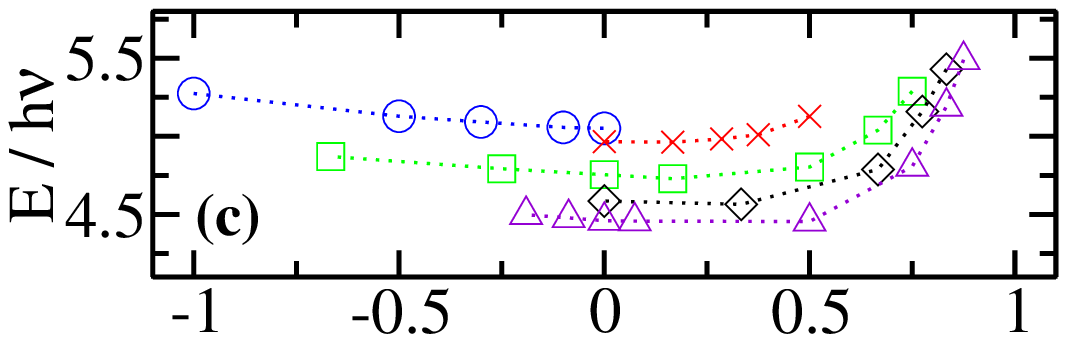}\\
\vspace*{-1.in}
\includegraphics[angle=270,width=90mm]{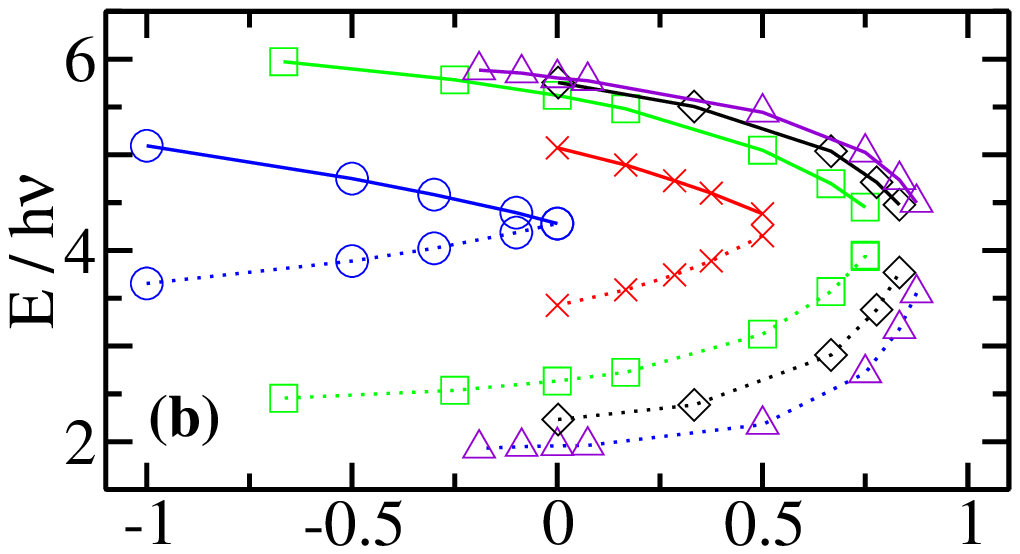}\\
\vspace*{-.5in}
\includegraphics[angle=270,width=90mm]{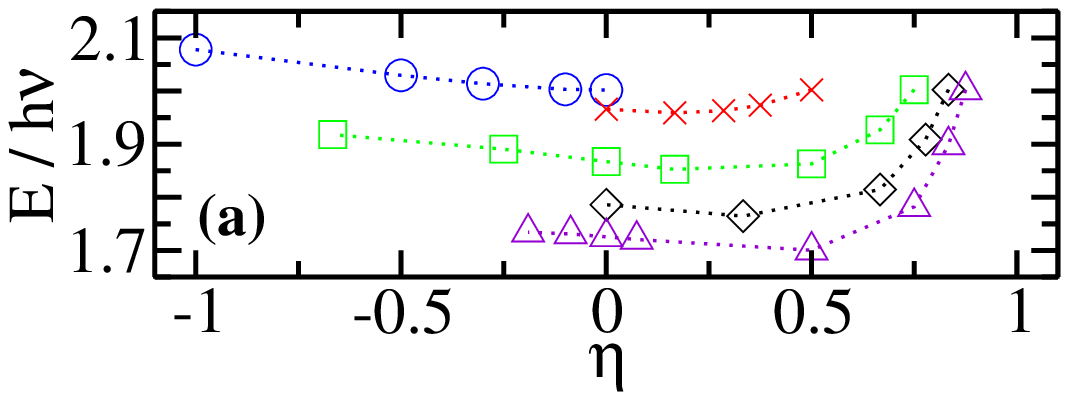}\\
\caption{
(Color online)
Ground state
energies in units of $\hbar \omega$
for two-component Fermi gases at unitarity as a function
of $\eta$ for (a) $N=2$, (b) $N=3$ and (c) $N=4$ for various 
mass ratios $\kappa$ for the square well potential with 
$R_0=0.01a_{ho,1}$.
The energies for $N=2$ are calculated by the DMC method and those 
for $N=3$ and 4 by the FN-DMC method.
The energies for $\kappa=1$, 2, 4, 6
and 8 are shown by circles, crosses, squares, diamonds and triangles,
respectively. For a given $\kappa$, symbols are
connected by lines to guide the eye:
Dotted lines are used for $N=2$, $N=3$ with two heavy 
atoms and one light atom, 
and $N=4$.
Solid lines are used for $N=3$ with two light 
atoms and 
one heavy atom.
For $\kappa=1$ and $N=3$, the dotted (solid) line 
connects the energies for systems in which two particles feel the 
smaller (larger) trapping frequency.
For each $\kappa$,
the energies for the largest $\eta$ value
considered correspond to $\omega_1=\omega_2$.
}
\label{fig1}
\end{figure}

Figure~\ref{fig1}(a)
shows the two-body energies $E_{1,1}$
calculated by the DMC method for the square well potential
with $R_0=0.01 a_{ho,1}$ and $1/|a_s|=0$
for $\kappa=1,2,4,6$ and 8 as a function of $\eta$.
For equal masses, 
Fig.~\ref{fig1}(a) shows the energies
for frequency ratios $\omega_2/\omega_1$ ranging from 
1/2 to 1.
For unequal masses, 
the ratio $\omega_2/\omega_1$ of trapping frequencies 
shown ranges from values
a bit smaller than $\kappa^{-1}$ to 1.
In units of $\hbar \omega$, the two-body energies for a fixed $\eta$
decrease with increasing mass ratio $\kappa$.
Furthermore, the minimum of the 
$E_{1,1}$ curves moves to larger 
$\eta$ as $\kappa$ increases.

To shed further light on the behavior of the two-body energies,
the fifth and sixth columns of Table~\ref{tab_n2} show the 
expectation value of $H_{cm}$, i.e., $E_{NL}$ with
$(NL)=(00)$, and the ground state
expectation value of $H_{rel}$ for the square well
potential with $R_0=0.01a_{ho,1}$.
The sum of these two expectation values coincides with the energy 
obtained for a single basis function
[namely, $\Phi_{NLM_L}\phi_{nlm_l}$
with $(NLM_Lnlm_l)=(000000)$]
in the diagonalization approach.
The difference between 
the fully converged energies (column~4 of Table~\ref{tab_n2})
and this sum is due to the coupling between the
center-of-mass and relative degrees of freedom. The
expectation value of $V_{coup}$ vanishes or is negative for all
two-body systems considered in this work and its magnitude increases
for a fixed $\kappa$ with increasing $\omega_1-\omega_2$.
For $\kappa=1$, the increase of 
$\langle H_{cm} +H_{rel}\rangle / (\hbar \omega)$
with increasing $\omega_1-\omega_2$
is larger than the decrease of $\langle V_{coup} \rangle / (\hbar \omega)$;
consequently, 
the equal-frequency system has the lowest energy.
For $\kappa=4$ and 8, 
$\langle H_{cm} \rangle /(\hbar \omega)$
first decreases with increasing $\omega_1-\omega_2$ and then increases
for $\omega_2/\omega_1 < 1/\kappa$, while the quantity
$\langle H_{rel} \rangle / (\hbar \omega)$ increases 
with increasing $\omega_1-\omega_2$ for all $\omega_2/\omega_1$.
It can be determined readily that the energy of the two-body system
at unitarity in the zero-range
limit without the coupling, $( \omega_{rel}/2 + 3 \omega_{cm}/2)/\omega$,
is minimal at $\eta \approx 0.43$ and $0.54$ for $\kappa=4$
and $8$, respectively. Since the absolute
value of
$\langle V_{coup} \rangle$ 
is fairly small compared to that of $\langle H_{cm}+H_{rel} \rangle$, 
the minimum of the energy $E_{1,1}/ ( \hbar \omega ) $ shifts
only slightly when the coupling term $V_{coup}$ is included
[see Fig.~\ref{fig1}(a)].

\begin{table}
\caption{
Selected expectation values, in units of $\hbar \omega$,
for the two-body system in the ground state with $N_1=N_2=1$ 
at unitarity
for $\kappa=1$, $4$ and 8
for various frequency ratios $\omega_2/ \omega_1$.
The energies in column 3 [superscript $(1)$]
are calculated for the square well potential
with $R_0=0.01a_{ho,1}$ 
using the DMC method; in this case,
the statistical uncertainty is in the last digit reported (or smaller).
The 
expectation values in columns 4-6
[superscript $(2)$] are calculated for the
square well potential
with $R_0=0.01a_{ho,1}$ using 
the diagonalization
scheme.
The energies in column 7 [superscript $(3)$]
are obtained by 
extrapolating the energies obtained by the 
diagonalization scheme for various $R_0$ to the $R_0 \rightarrow 0$ limit;
the extrapolation error is estimated to be at most $0.001\hbar \omega$.
}
\begin{ruledtabular}
\begin{tabular}{cc|l|lll|l}
  $\kappa$ & $\omega_2/\omega_1$ & $E_{1,1}$$^{(1)}$ & $E_{1,1}$$^{(2)}$ & $\langle H_{cm} \rangle$$^{(2)}$ & $\langle H_{rel}\rangle$$^{(2)}$ & $E_{1,1}$$^{(3)}$ \\
\hline
1 & 1     & 2.003 & 2.002 & 1.500 & 0.502 & 2.000 \\
  & 10/11 & 2.003 & 2.004 & 1.502 & 0.503 & 2.002 \\
  & 10/13 & 2.014 & 2.014 & 1.513 & 0.506 & 2.012 \\
  & 2/3   & 2.030 & 2.029 & 1.530 & 0.512 & 2.028 \\ \hline
4 & 1     & 2.002 & 2.003 & 1.500 & 0.503 & 2.000 \\
  & 3/4   & 1.927 & 1.927 & 1.382 & 0.549 & 1.924 \\
  & 1/2   & 1.863 & 1.862 & 1.265 & 0.618 & 1.859 \\
% & 0.3 & 1.853 & & \\
  & 1/4   & 1.867 & 1.866 & 1.200 & 0.725 & 1.863 \\
% & 1/5 & 1.891 & &  & \\
  & 3/20  & 1.918 & 1.921 & 1.218 & 0.784 & 1.918 \\ \hline
8 & 1     & 2.003 & 2.003 & 1.500 & 0.503 & 2.000 \\
  & 3/4   & 1.898 & 1.898 & 1.340 & 0.560 & 1.895 \\
  & 1/2   & 1.782 & 1.783 & 1.155 & 0.642 & 1.780 \\
  & 1/4   & 1.700 & 1.701 & 0.980 & 0.761 & 1.697 \\
% & 0.135 & 1.721 & 1.719 & \\
 & 1/8    & 1.726 & 1.724 & 0.943 & 0.843 & 1.720 \\
% & 0.115 & 1.731 & 1.730 & \\
% & 0.105 & 1.724 & 1.737 & 0.944 & 0.858 & \\
\end{tabular}
\end{ruledtabular}
\label{tab_n2}
\end{table}

The two-body system with unequal frequencies has been discussed
previously by a number of groups. 
The energies of the lowest-lying gas-like states and the most weakly-bound 
molecular states of the 
trapped $^{40}$K-$^{87}$Rb dimer have, e.g., been measured experimentally 
and been determined theoretically as part of a project
on  Fermi-Bose mixtures in a lattice~\cite{deur07}.
Also, the effect of the coupling between the center-of-mass
and relative motions has been investigated in the context
of confinement-induced 
resonances~\cite{stoc03,kim06,mele07}. 
Our main focus lies in extending the two-body study presented above 
to larger unequal-frequency systems with three, four or more
particles per lattice site. 
While an increasing body of literature exists for larger
equal-frequency systems, the regime 
where the center-of-mass motion does not decouple has,
to the best of our knowledge, received only 
little attention for larger systems,
despite its immediate relevance to ongoing experiments.

Figures~\ref{fig1}(b) and (c)
show the FN-DMC energies, in units
of $\hbar \omega$, for two-component Fermi gases with $N=3$ and 4
as a function of $\eta$ for various mass ratios $\kappa$.
The overall behavior of the $N=4$ energies is similar to that of the $N=2$
energies.
For a given $\eta$, the $N=4$ energies 
decrease 
with increasing $\kappa$.
For $\kappa=1$, the energy
$E_{2,2}$ is minimal for $\eta \approx 0$ and increases with
increasing $|\eta|$. As $\kappa$ increases, 
$E_{2,2}$ is minimal 
for positive $\eta$ values. 
As discussed above,
the $N=2$ eigenenergies for equal frequencies
at unitarity approach
$2 \hbar \omega$ 
in the zero-range limit for all $\kappa$.
For $N=4$, in contrast, the energies 
at unitarity depend on the mass ratio even when the trapping frequencies
coincide~\cite{stec08}.

The $N=3$ energies at unitarity behave
qualitatively different from the $N=2$ and 4
energies. 
For systems with one spare heavy atom
[symbols connected
by dotted lines in Fig.~\ref{fig1}(b)], the energy for a given $\kappa$
decreases with decreasing $\eta$. 
For systems with 
one spare light atom, the behavior is reversed, i.e., the
energy increases with decreasing $\eta$. This behavior is
similar to that of the coefficients
$C_{1,2}$ and $C_{2,1}$ (see Fig.~\ref{fig_pert}), 
and can qualitatively, as in the perturbative
regime, be explained
in terms of the overlap of the densities of the two components.

Table~\ref{tab_n3to6} summarizes selected FN-DMC energies for $N=3-6$
at unitarity calculated for the square well potential with $R_0=0.01 a_{ho,1}$.
The $N=4$ energies for $\eta=0$ and $\kappa=4$ and 8  
are slightly lower than those reported in
Ref.~\cite{stec07b}. This is due to the fact that 
the $\kappa >1$ calculations of Ref.~\cite{stec07b} for even $N$ 
were restricted 
to the guiding function $\psi_{T1}$, while this paper 
also considers the guiding function $\psi_{T2}$, whose nodal surface
coincides with that of the non-interacting system. For $N=4$ and $\eta=0$,
the energies obtained using $\psi_{T2}$ are lower than those obtained 
using
$\psi_{T1}$.
Table~\ref{tab_n3to6} 
also indicates the total angular momentum $L_{tot}$ of the
lowest energy state obtained by the FN-DMC method.
For $N=4$ and 6 (no superscript in Table~\ref{tab_n3to6}), 
we obtain the lowest energy for a guiding function
with $L_{tot}=0$
for all mass and trapping frequency ratios considered.
For $N=3$, in contrast, the total angular momentum of the lowest energy
state depends on the system considered and is either 0 or 1.
For $N=5$, the lowest energy state found by the FN-DMC method has $L_{tot}=1$.

\begin{figure}
\includegraphics[angle=270,width=90mm]{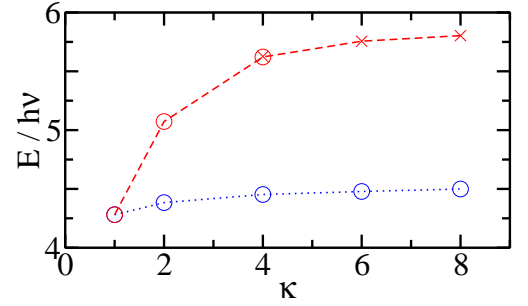}
\caption{
(Color online)
Ground state energies
for $N=3$ 
(two light atoms and one heavy atom)
in units of $\hbar \omega$ as a function of $\kappa$ 
at unitarity; the energies are calculated 
by the FN-DMC method for the square well potential
with $R_0=0.01a_{ho,1}$.
The symbols connected by dotted and dashed lines are calculated 
for equal trapping frequencies and equal trapping lengths, respectively.
The energies shown by circles and crosses correspond to states
with total angular momentum $L_{tot}=1$ and 0, respectively.
}
\label{fig_n3ang}
\end{figure}

To investigate the ``angular momentum crossover'' of the three-particle system 
with two light atoms and one heavy
atom in more detail, Fig.~\ref{fig_n3ang}
shows the FN-DMC energies for equal frequencies 
(symbols connected by a dotted
line)
and for equal trapping lengths (symbols connected by
a dashed line)
as a function of the mass ratio $\kappa$.
Energies of states with $L_{tot}=1$ and 0 are shown
by circles and crosses,
respectively.
For equal trapping frequencies, the angular momentum of the lowest
energy state is $L_{tot}=1$ for all
mass ratios considered.  
For equal trapping lengths, in contrast,
$L_{tot}$ changes from 1 to 0 at
$\kappa  \approx 4$.
This angular momentum crossover has also been observed in
calculations that employ the correlated Gaussian (CG) 
approach~\cite{javier_cg}.

\begin{table}
\caption{\label{tab_n3to6} 
Selected FN-DMC energies
$E_{N_1,N_2}$, in units of $\hbar \omega$, for $N=3-6$ ($|N_1-N_2| \le 1$)
at unitarity
for $\kappa=4$ and 8.
The FN-DMC energies are uncertain in the last digit
reported.
The guiding functions used to obtain the
energies marked by a superscript ``$*$'' have total angular momentum
$L_{tot}=1$ and those not marked by a superscript have $L_{tot}=0$.
The superscript ``$?$'' marks systems for which
guiding functions with $L_{tot}=0$ and $1$ result in energies that  
are indistinguishable within the statistical uncertainties.
For comparison, the CG approach results 
in 
$E_{2,1}=5.67 \hbar \omega$,
$E_{1,2}=1.96 \hbar \omega$ and
$E_{2,2}=4.45 \hbar \omega$
for $\kappa=8$, $\omega_2/ \omega_1=1/8$
and a range comparable to that employed in the FN-DMC 
calculations~\protect\cite{javier_cg};
as in the equal-frequency case~\protect\cite{stec07b,stec07c,stec08}, 
the FN-DMC energies compare
favorably with the energies calculated by the CG approach.
}
\begin{ruledtabular}
\begin{tabular}{cc|llllll}
  $\kappa$ & $\omega_2/\omega_1$ & $E_{2,1}$ & $E_{1,2}$ & $E_{2,2}$ & $E_{3,2}$ & $E_{2,3}$  & $E_{3,3}$\\
\hline
4 & 1    & 4.45$^{*}$ & 3.94$^{*}$ & 5.29 & 7.99$^{*}$ & 7.44$^{*}$ & 9.51 \\ 
  & 3/4  & 4.70$^{*}$ & 3.57$^{*}$ & 5.04 & 8.08$^{*}$ & 6.88$^{*}$ & 8.94 \\ 
  & 1/2  & 5.05$^{*}$ & 3.13$^{*}$ & 4.80 & 8.26$^{*}$ & 6.27$^{*}$ & 8.44 \\ 
  & 1/4  & 5.62$^{?}$ & 2.63$^{*}$ & 4.76 & 8.81$^{*}$ & 5.72$^{*}$ & 8.24 \\ 
  & 3/20 & 5.97$^{?}$ & 2.45$^{*}$ & 4.87 & 9.24$^{*}$ & 5.62$^{*}$ & 8.43 \\ \hline
8 & 1    & 4.50$^*$ & 3.55$^{*}$ & 5.49 & 8.19$^{*}$ & 7.12$^{*}$ & 9.59 \\ 
  & 3/4  & 4.74     & 3.18$^{*}$ & 5.19 & 8.25$^{*}$ & 6.64$^{*}$ & 9.21 \\ 
  & 1/2  & 5.03     & 2.72$^{*}$ & 4.81 & 8.31$^{*}$ & 5.90$^{*}$ & 8.78 \\ 
  & 1/4  & 5.44     & 2.18$^{*}$ & 4.46 & 8.56$^{*}$ & 5.08$^{*}$ & 7.90 \\ 
  & 1/8  & 5.80     & 1.96$^{*}$ & 4.46 & 8.96$^{*}$ & 4.83$^{*}$ & 7.82 \\ 
\end{tabular}
\end{ruledtabular}
\end{table}

\begin{figure}
\includegraphics[angle=270,width=90mm]{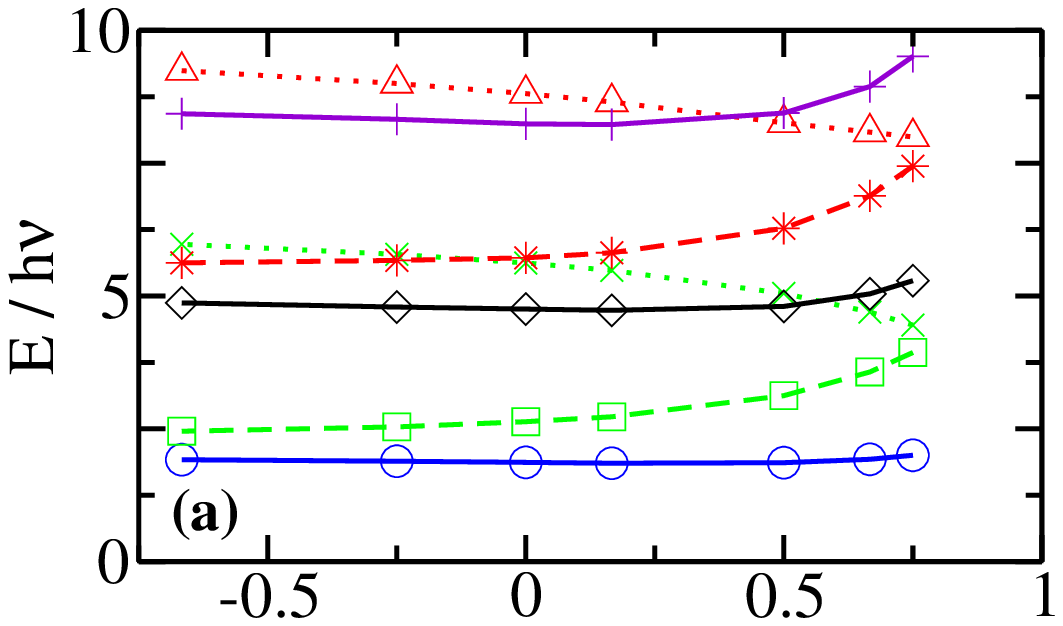}
\vspace*{-0.1in}
\includegraphics[angle=270,width=90mm]{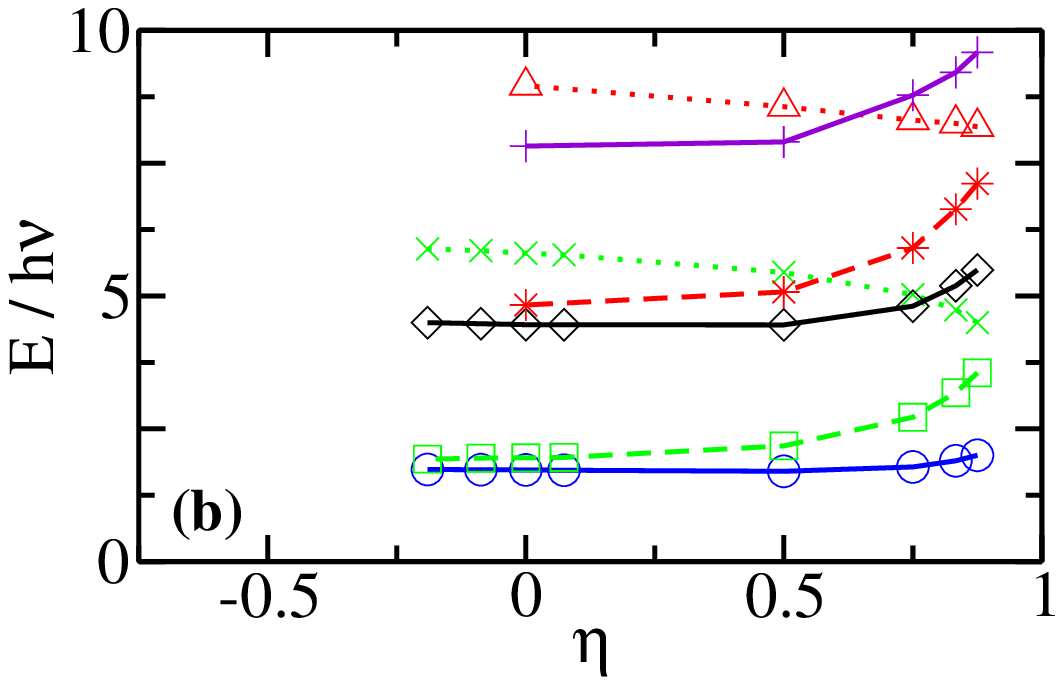}
\caption{
(Color online)
Ground state energies at unitarity for (a) $\kappa=4$ 
and (b) $\kappa=8$ as a funcion of $\eta$ for
$N=2$ (circles), $N=3$ (squares and crosses), $N=4$ (diamonds),
$N=5$ (asterisks and triangles) and $N=6$
(pluses). Energies for even $N$ systems are connected by solid lines,
and those for odd $N$ systems with a spare heavy and a spare light particle
by dashed and dotted lines, respectively.
The calculations are performed using the DMC ($N=2$)
and the FN-DMC ($N=3-6$) method for the square well potential with 
$R_0=0.01a_{ho,1}$.
}
\label{fig_kappa4}
\end{figure}

Figure~\ref{fig_kappa4}
shows the energies for $N=2-6$ with $\kappa=4$ 
and $8$ 
as a function of $\eta$.
The energies for $N=2$, $4$ and $6$
behave similarly
as a function of $\eta$, with the $N=2$ energies being  nearly
constant [see also Fig.~\ref{fig1}(a)]
and the $N=4$ and 6 energies showing a stronger decrease
than the $N=2$ energies
as $\eta$ decreases from $3/4$ to $\approx 0.4$ for $\kappa=4$ and from
$7/8$ to $\approx 0.5$ for $\kappa=8$
[see also the discussion in the 
context of Figs.~\ref{fig1}(b) and (c)].
Furthermore, the energies
$E_{2,1}$ and $E_{3,2}$, and the energies
$E_{1,2}$ and $E_{2,3}$
show a similar overall behavior.
The ordering of the energy levels for equal trapping frequencies
(for $\kappa=4$, e.g., this corresponds to $\eta=3/4$) 
is, from bottom to top, 
$E_{1,1}$, $E_{1,2}$, $E_{2,1}$, $E_{2,2}$, 
$E_{2,3}$, $E_{3,2}$ and $E_{3,3}$.
This ordering changes as $\eta$ decreases;
for $\kappa=4$, e.g., $E_{2,1}$ becomes larger than
$E_{2,2}$ at $\eta \approx 0.6$, and also larger than $E_{2,3}$ 
at $\eta \approx -0.2$.
Similarly, $E_{3,2}$ becomes larger than $E_{3,3}$ at $\eta \approx 0.4$.

\begin{figure}
\includegraphics[angle=270,width=90mm]{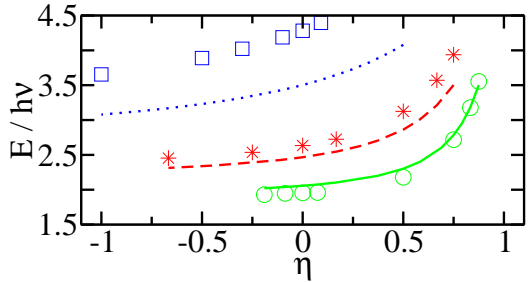}
\caption{
(Color online)
Comparison of the energies at unitarity
for the (3+0)-configuration calculated by the
FN-DMC method (symbols) and
those for the (2+1)-configuration calculated by the diagonalization
scheme (lines) as a function of $\eta$. 
The energies for the (3+0)-configuration 
are shown by squares, asterisks and circles for $\kappa=1$, 4 and 8,
respectively, while those for the (2+1)-configuration 
are shown by a dotted, dashed and solid line for $\kappa=1$, 4 and 8, respectively.
For $\kappa=1$, two particles feel the larger (smaller) angular
trapping frequency for $\eta>0$ ($\eta<0$);
for $\kappa>1$, the systems considered
consist of two heavy fermions and one light fermion.
}
\label{fig_lattice}
\end{figure}

The small $N$ energies 
can be combined to predict the 
energetically most favorable configuration of an optical lattice with
small tunneling amplitude, approximately harmonic lattice sites,
twice as many particles of one species
than of the other and a filling factor equal to or
smaller than 3/2. For equal masses and equal trapping frequencies, 
it has been shown previously for all $a_s$~\cite{kest07,stet07,stec08} 
that it is energetically more favorable for one spin-up and one spin-down
atom to occupy one lattice site and for the second spin-down
atom to occupy a different site 
[we refer to this as the ``(2+1)-configuration''] than for the one spin-up
and the two spin-down fermions
to occupy the same lattice site [we refer to this
as the ``(3+0)-configuration'']. Figure~\ref{fig_lattice}
extends this analysis to trap- and mass-imbalanced systems
at unitarity. A dotted line shows the energy of the (2+1)-configuration
for $\kappa=1$ at unitarity as a function of $\eta$
[the particle in the singly-occupied lattice site 
feels the larger (smaller) frequency
for $\eta>0$ ($\eta < 0$)], while squares show the energy of the 
(3+0)-configuration. No crossing between these two curves is observed, with
the energy of the (3+0)-configuration being larger than the energy of the 
(2+1)-configuration.
We find a similar behavior for $\kappa=4$: For the
frequency ratios considered, 
the energy increase due to placing two 
like fermions in the same lattice site is so large
that it is more favorable to place the like fermion
in a separate lattice site instead (and thereby
``loosing'' the energy decrease due to having
two like fermions interact with the unlike fermion through an attractive
two-body potential).
For $\kappa=8$, however, the behavior is different: Figure~\ref{fig_lattice}
shows that the  (2+1)-configuration (solid line) is energetically favorable
for larger $\eta$ and the (3+0)-configuration (circles) is energetically
favorable for smaller $\eta$. The crossover is predicted to occur at $\eta \approx 0.8$.
This suggests that it might be possible to introduce a 
macroscopic phase
transition of an 
optical lattice
system by changing the trapping frequency felt
by one of the species if the mass ratio is sufficiently large but not
necessarily so large that three-body bound states with molecular character exist.

We now discuss 
the odd-even oscillations of larger trapped unequal-mass systems
at unitarity. In particular,
we focus on systems with $\eta=0$, i.e., on systems for which the two 
oscillator lengths $a_{ho,1}$ and $a_{ho,2}$ coincide.
Equal-mass systems with both even and odd $N$ have already been discussed in 
Refs.~\cite{blum07,stec08} and unequal-mass systems with 
even $N$ 
in Ref.~\cite{stec07b}.
Results for unequal-mass systems with odd $N$ ($N \ge 5$), in contrast, have not been
presented before.
For even $N$, $N \ge 6$, we obtain the lowest FN-DMC energy for 
$\kappa = 1$, 4 and 8
for the guiding
function $\psi_{T1}$ (see Sec.~\ref{sec_num}), whose nodal surface is
constructed by anti-symmetrizing a two-body pair function.
For odd $N$, we find that the guiding function that results in the lowest
FN-DMC energy
depends not only on $N$ but also on $\kappa$:
For $\kappa=1$, the guiding function
$\psi_{T2}$ results in the lowest FN-DMC energy for 
$N \le 9$, $\psi_{T3}$ for $N=11$ and 
$\psi_{T1}$ 
for 
$N \ge 13$.
For $\kappa=4$ and 8 systems with a spare light atom,
the guiding functions $\psi_{T2}$ and $\psi_{T1}$
result in the lowest FN-DMC energy for $N \le 7$ and $N \ge 9 $, respectively.
For $\kappa=4$ and 8 systems with a spare heavy atom, in contrast,
the guiding functions
$\psi_{T2}$ and $\psi_{T3}$ result in the lowest FN-DMC energy
for $N \le 9$ and $N \ge 11$, respectively.
The density profiles reveal that the spare particle
of the odd-$N$ systems 
with $N \gtrsim 11$ and $\kappa=1$  is
located predominantly
near the edge of the cloud~\cite{blum07,stec08}.
Thus, one may consider the core region as ``fully paired''
and the edge region as ``partially paired''.
For $\kappa>1$ systems with a spare heavy fermion, 
the pairing function $\psi_{T1}$ results in a higher energy 
than $\psi_{T3}$ for $N \ge 9$, suggesting that the nodal surface $\psi_{T3}$ 
allows for a higher probability of three fermions to be in close proximity
than the nodal surface of $\psi_{T1}$
(recall, for sufficiently
large $\kappa$ three-body bound states with one quantum of angular momentum exist).

\begin{figure}
\includegraphics[angle=270,width=90mm]{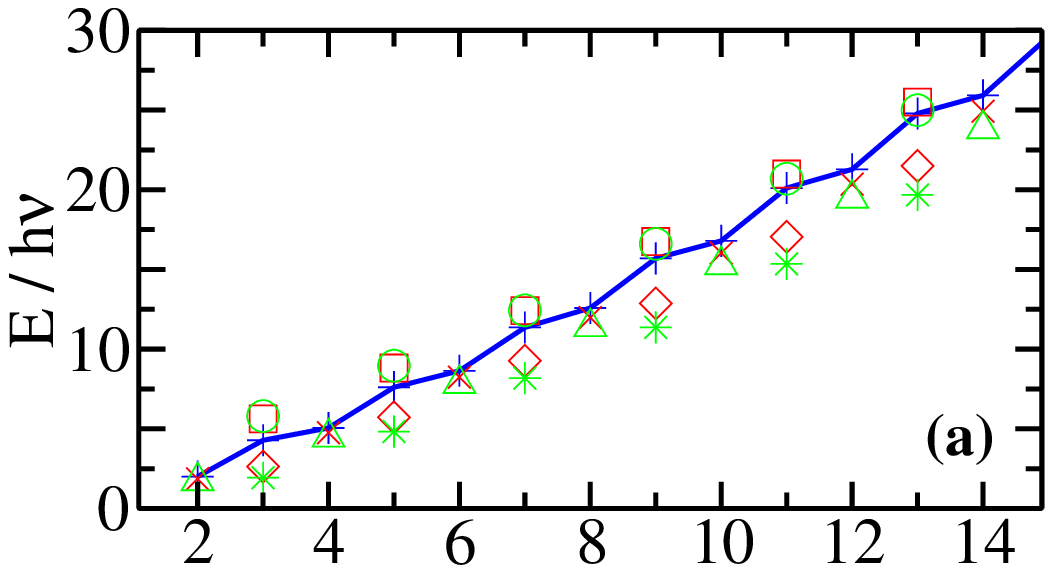}\\
\includegraphics[angle=270,width=90mm]{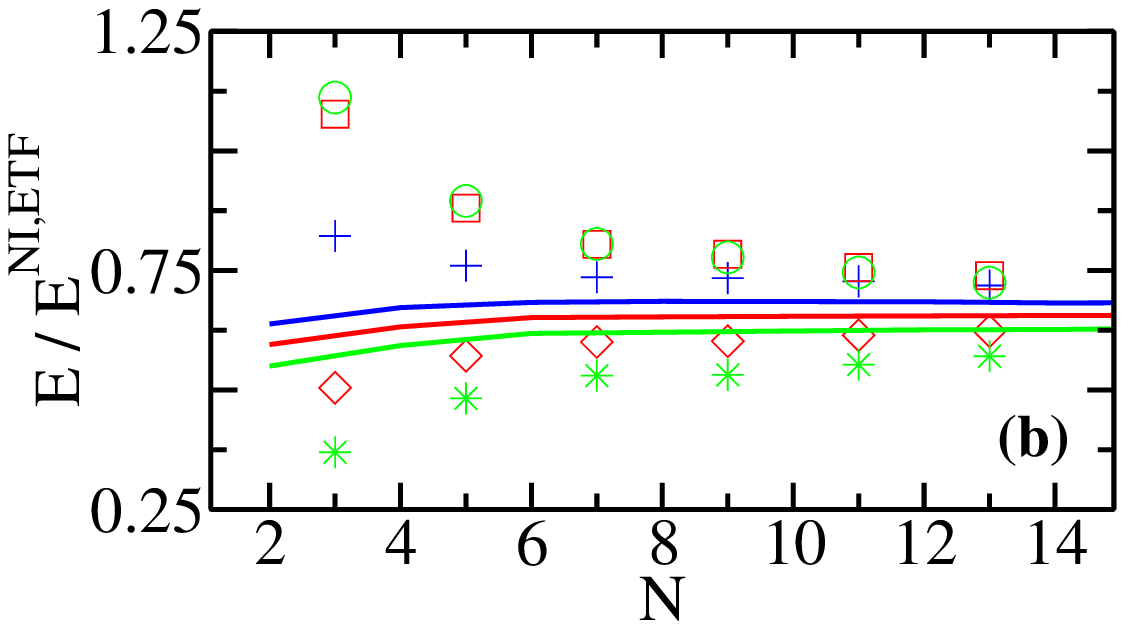}
\caption{
(Color online)
Energies in units of (a) $\hbar \omega$ 
and (b) $E^{NI,ETF}$ at unitarity for $\eta=0$
as a function
of $N$ for $\kappa=1$, 4 and 8.
The energies are calculated for the square well potential
with $R_0=0.01 a_{ho,1}$ using the DMC ($N=2$) and FN-DMC ($N=3-14$) methods.
(a) $\kappa=1$ (pluses connected by a solid line), $\kappa=4$ 
[crosses for even $N$,
and diamonds (spare heavy atom) and 
squares (spare light atom) for odd $N$],
and $\kappa=8$
[triangles for even $N$,
and asterisks (spare heavy atom) and 
circles (spare light atom) for odd $N$].
(b) $\kappa=1$ (uppermost
solid line for even $N$ and pluses for odd $N$), $\kappa=4$ 
[middle solid line for even $N$,
and diamonds (spare heavy atom) and 
squares (spare light atom) for odd $N$],
and $\kappa=8$
[lowermost solid line for even $N$,
and asterisks (spare heavy atom) and 
circles (spare light atom) for odd $N$].
}
\label{fig_largen1}
\end{figure}

Figures~\ref{fig_largen1}(a) and (b) show 
the energies for systems with up to $N=14$ atoms
with $\kappa=1$, 4 and 8. 
The even-$N$ energies,
shown in units of $\hbar \omega$ in
Fig.~\ref{fig_largen1}(a), decrease for a given $N$ with increasing 
$\kappa$
[for $N=2$ and 4, see also Figs.~\ref{fig1}(a)
and (c)].
The energies for systems 
with a spare heavy
particle (diamonds and asterisks for $\kappa=4$ and
8, respectively) are notably smaller than the corresponding 
odd $N$ energies for $\kappa=1$.
The energies for systems with a spare light particle
(squares and circles for $\kappa=4$ and 8, respectively),
in contrast, are higher  than the corresponding
energies for $\kappa=1$ for small $N$ and nearly
coincide with
the corresponding energies for $\kappa=1$ for
larger $N$.
Further optimization of the nodal surface of the
guiding functions employed in the FN-DMC calculations 
may result in tighter upper bounds for the energies;
a more detailed investigation of larger odd-$N$ systems with $\kappa>1$
is relegated to the future.

To see the odd-even oscillations more clearly, Fig.~\ref{fig_largen1}(b)
scales the energies from panel (a)
by the ``smoothed'' extended Thomas-Fermi
energies $E^{NI,ETF}$ of the non-interacting system~\cite{brac97}, 
\begin{eqnarray}
\label{eq_etf}
E^{NI,ETF}= \hbar \omega \frac{(3N)^{4/3}}{4} 
\left[ 1+\frac{(3N)^{-2/3}}{2} \right].
\end{eqnarray}
For $\kappa=1$, the scaled energies follow two distinct curves;
the curve for odd $N$ (pluses) is higher than that for even $N$
(topmost solid line),
reflecting the non-vanishing excitation gap 
at unitarity (see below and
Refs.~\cite{blum07,stec08,chan07}).
For fixed $N$, 
the scaled even-$N$ energies decrease
with increasing $\kappa$~\cite{stec07b}.
The scaled energies for odd-$N$ systems with
one spare heavy atom are lower 
than the corresponding scaled energies
for systems with $N-1$ fermions and the same $\kappa$.
The scaled energies for odd-$N$ systems with one spare
light particle, in contrast,
are higher than the corresponding scaled energies for systems with
$N-1$ fermions and the same $\kappa$.

\begin{figure}
\includegraphics[angle=270,width=90mm]{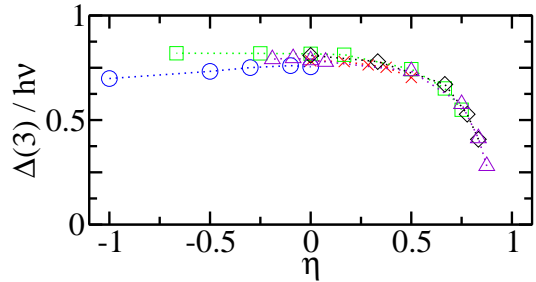}
\caption{
(Color online)
Excitation gap $\Delta(3)$ in units of $\hbar \omega$
as a function of $\eta$
for two-component Fermi gases at unitarity for
various $\kappa$. The
$\Delta(3)$, calculated using the 
DMC ($N=2$) and FN-DMC ($N=3$ and 4) energies,
are shown by circles, crosses, squares, diamonds and triangles
 for $\kappa=1$, 2, 4, 6
and 8, 
respectively. Dotted lines connect symbols for a fixed $\kappa$ 
to guide the eye.
}
\label{fig_gap3}
\end{figure}

Next, we
combine our energies for even and odd $N$
to determine the excitation gap $\Delta(N)$, Eq.~(\ref{eq_gap}),  
at unitarity. 
Figure~\ref{fig_gap3} shows $\Delta(N)$ for $N=3$
as a function of $\eta$ for various $\kappa$.
Although 
the FN-DMC energies themselves provide an upper bound to the 
exact eigenenergies, the excitation gap
is not variational.
Figure~\ref{fig_gap3} shows that the excitation gap $\Delta(3)$ 
for $\kappa=1$ through 8 collapse---for the systems
considered---approximately 
to a single curve for $\eta \gtrsim 0$; 
$\Delta(3)$ is maximal around $\eta=0$ and decreases
with increasing $\eta$. While more detailed calculations may 
reveal the dependence of $\Delta(3)$ on $\kappa$ for a given
$\eta$ in more detail, our calculations suggest that
$\Delta(3)$ is determined predominantly by $\eta$ and only secondarily
by $\kappa$. This is in contrast to the energies
themselves (see, e.g., 
Fig.~\ref{fig1}) and can be attributed at least partially 
to the
fact that the energies of the two odd-$N$ systems
(that with a spare heavy and that with a spare light
particle)
enter into
$\Delta(3)$ as an averaged quantity.

\begin{figure}
\includegraphics[angle=270,width=90mm]{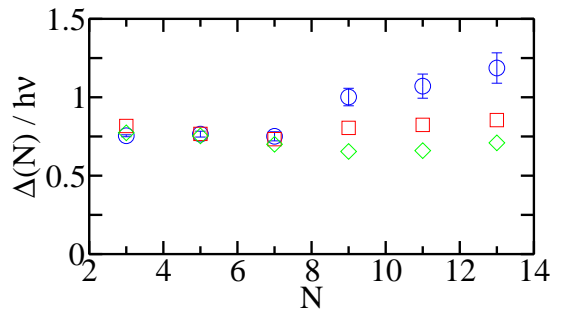}
\caption{
(Color online)
Excitation gap $\Delta(N)$
in  units of $\hbar \omega$ 
at unitarity for $\eta=0$ (i.e., for $m_1 \omega_1 = m_2 \omega_2$)
as a function
of $N$ for $\kappa=1$ (circles with errorbars), $\kappa=4$ 
(squares),
and $\kappa=8$
(diamonds). 
The errorbars of $\Delta(N)$ for $\kappa=4$ and 8, which are not
shown to enhance the clarity of the figure, are a bit larger than those for
$\kappa=1$.
The excitation gap 
$\Delta(N)$ is calculated from the energies
shown in Fig.~\protect\ref{fig_largen1}.
}
\label{fig_largen2}
\end{figure}

For larger $N$, we
determine the excitation gap
$\Delta(N)$
at unitarity for $\kappa=1$, 4 and 8
for systems with equal trapping lengths, i.e., for
$a_{ho,1}=a_{ho,2}$.
Figure~\ref{fig_largen2} shows that $\Delta(N)$, 
expressed in units of $\hbar \omega$, is
nearly constant ($\approx 0.75 \hbar \omega$)
for $N=3, 5$ and 7 for all mass ratios considered
(for $N=3$, see also Fig.~\ref{fig_gap3}).
For $N \ge 9$, the excitation gap $\Delta(N)$
is largest for $\kappa=1$ and smallest for $\kappa=8$.
Despite the fairly large uncertainties of the excitation gap
(see caption of Fig.~\ref{fig_largen2}), we are
quite confident that the excitation gap
does indeed decrease with increasing $\kappa$ but fixed $N$ ($N \gtrsim 9$).
This is a direct consequence of the decrease of the energies
for odd-$N$ systems with a spare heavy particle  with increasing 
$\kappa$.

\subsection{Small positive $s$-wave scattering length}
\label{sec_positive}
This section discusses the behavior of two-component
Fermi gases with unequal masses and unequal frequencies
in the BEC regime where the $s$-wave scattering length $a_s$
is small and positive.
In this regime, the behavior of the molecular 
Bose gas is expected to be governed by the
dimer-dimer scattering length $a_{dd}$~\cite{petr04,petr05,petr05a}.
For equal-frequency systems, the four-fermion spectrum 
has been
compared with that of two bosons~\cite{stec07b,stec08}, 
validating the ``dimer picture''.
This section extends the previous analysis to two-component Fermi
gases with unequal frequencies.

\begin{figure}
\includegraphics[angle=270,width=90mm]{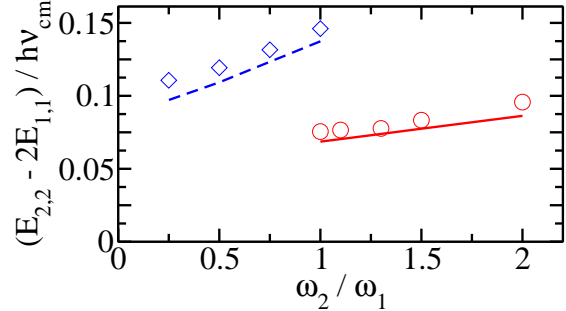}
\caption{
(Color online)
Interaction shifts in units
of $\hbar \omega_{cm}$ 
for four-particle system 
in the BEC regime.
Circles and diamonds show the energy difference $E_{2,2}-2 E_{1,1}$ 
for $\kappa=1$ and 4, respectively;
the uncertainty of $(E_{2,2}-2E_{1,1})/(\hbar \omega_{cm})$
is about $0.01 \hbar \omega_{cm}$. 
The energies
$E_{2,2}$ and $E_{1,1}$
are calculated by the FN-DMC and diagonalization
approaches, respectively, for the square
well potential with $R_0=0.01a_{ho,1}$
and $a_s = 0.1a_{ho,1}$.
For comparison, solid and dashed lines show the interaction
shift $E_{dd}$ determined perturbatively, Eq.~(\ref{eq_edd}),
for $\kappa=1$ and 4, respectively.
The quantity $a_s/a_{ho,cm}$ increases with increasing
$\omega_2/\omega_1$, which explains the increase of the interaction shift
with increasing $\omega_2/\omega_1$.
}
\label{fig_bec}
\end{figure}

We determine the two-body energy $E_{1,1}$ for unequal
frequency systems using the diagonalization scheme.
For a sufficiently small atom-atom scattering length, 
we find that the two-body
energy is to a very good approximation
given by the sum of the expectation values of $H_{cm}$ 
and $H_{rel}$ (the expectation value of the
coupling term $V_{coup}$ is smaller than $10^{-4} \hbar \omega$
for the systems considered in Fig.~\ref{fig_bec}).
Thus, an approximate but highly accurate expression
for the ground state energy of the two-body system
on the BEC side
reads
\begin{eqnarray}
E_{1,1} \approx \langle H_{rel} \rangle_{000} + \frac{3}{2} \hbar \omega_{cm}.
\end{eqnarray}
Assuming diatomic molecules form, the lowest energy of the
four-particle system with $N_1=N_2=2$ and small $a_s$
can be written as
\begin{eqnarray} 
E_{2,2}
\approx 2 \langle H_{rel} \rangle_{000} + \frac{3}{2} 
\hbar \omega_{cm} + E_{rel,boson},
\end{eqnarray}
where
the first term on the right hand side is the internal energy
of the two bosonic molecules, the second term on the right hand side
is the center-of-mass energy
of the two-boson system and the third term on the right
hand side is the relative energy of the
two-boson system.
We rewrite the latter as
\begin{eqnarray}E_{rel,boson} = \frac{3}{2} \hbar \omega_{cm} + E_{dd},
\end{eqnarray}
and evaluate 
the ``dimer-dimer interaction shift'' $E_{dd}$ 
in first order perturbation theory (assuming a Fermi contact 
potential)~\cite{ferm34,busc98},
\begin{eqnarray}
\label{eq_edd}
E_{dd} \approx \sqrt{\frac{2}{\pi}} \, \frac{a_{dd}}{a_{ho,cm}}
\, \hbar \omega_{cm}.
\end{eqnarray}
Here, $a_{ho,cm}$ denotes the oscillator length associated
with $\omega_{cm}$, i.e., $a_{ho,cm} = \sqrt{\hbar/(M \omega_{cm})}$.
It follows that the energy difference $E_{2,2}-2 E_{1,1}$
should be given by the interaction shift $E_{dd}$.
The reasoning outlined here for four fermions can be extended to larger 
systems: The trapping frequency $\omega_{rel}$ determines---together 
with the atom-atom scattering length $a_s$---the
internal binding energy of the molecules while the trapping frequency
$\omega_{cm}$ determines---together with the dimer-dimer
scattering length $a_{dd}$---the properties of the composite boson
system.

Figure~\ref{fig_bec} shows the difference between the 
FN-DMC energy $E_{2,2}$ and twice the two-body energy $E_{1,1}$
as a function of the frequency ratio $\omega_2/\omega_1$
for $\kappa=1$ (circles) and $\kappa=4$ (diamonds).
For comparison, solid and dashed lines show the interaction
shift $E_{dd}$, Eq.~(\ref{eq_edd}),
for $\kappa=1$ (using $a_{dd}=0.608a_s$~\cite{petr04,astr04c,stec07b,stec08}) 
and $\kappa=4$ (using $a_{dd}=0.77a_s$~\cite{petr05,stec07b,stec08}), 
respectively. The numerically determined energy differences
are a bit larger than the perturbative result, which is
in agreement with the fact that the FN-DMC energy $E_{2,2}$ provides
an upper bound to the true eigenenergy~\cite{reyn82}. The
agreement between
the interaction shift obtained by
solving the full four-body Schr\"odinger
equation and by treating the weakly-interacting two-boson system
perturbatively is similarly good for all trapping frequencies considered.
This confirms that the relevant ``boson frequency'' is indeed,
as has been argued previously by others~\cite{iski08,orso07}, given
by $\omega_{cm}$. 

\section{Conclusion}
\label{sec_conclusion}
This paper determines the ground state energies of two-component Fermi systems
under external harmonic confinement with unequal masses
and unequal frequencies.
We considered the weakly-interacting, small $|a_s|$ regime with both
positive and negative scattering lengths $a_s$ as well as the strongly-interacting 
unitary regime where the $s$-wave scattering length diverges. 
In all three regimes, we identified convenient energy and length units.
In the weakly-attractive regime, we treated the atomic Fermi gas perturbatively. 
In the unitary regime, we determined the eigenenergies numerically. 
In the weakly-repulsive regime, 
we compared numerical results with those obtained by treating
the weakly-repulsive molecular Bose gas perturbatively.

The calculations at unitarity are performed for a short-range potential with
small range; the resulting energies are estimated to be quite close to the 
zero-range
limit. We determine the energies as a function of both the ratio
between the masses of the two species   and the ratio between the trapping 
frequencies felt by the two species. 
The small $N$ results presented cover a wide range of mass and
frequency ratios and
can easily be extrapolated to experimentally
relevant parameter combinations (such as $^6$Li-$^{40}$K mixtures).
Our FN-DMC energies provide an upper bound to the true ground state energy
of mass- and trap-imbalanced two-component Fermi systems
and may be serve as a benchmark for other approaches.

DB is grateful to J. von Stecher for calculating
the CG energies reported in 
the caption of Table~\ref{tab_n3to6}
and for discussions during the
initial stage of this work. 
DB also  gratefully acknowledges support by the NSF through
grant PHY-0555316.

%\bibliography{lit}
%\bibliographystyle{prsty}

\end{document}